\documentclass[12pt,english]{article}
\usepackage[T1]{fontenc}
\usepackage[latin9]{inputenc}
\usepackage[letterpaper]{geometry}
\usepackage{float}
\usepackage{amsmath}
\usepackage{graphicx}
\usepackage{setspace}
\usepackage{amssymb}
\usepackage{babel}

\begin{document}

\title{Entropy of cosmological black holes and generalized second law in
phantom energy-dominated universe }

\author{Khireddine Nouicer%
\thanks{E-mail:khnouicer@univ-jijel.dz%
}}

\date{}
\maketitle
\begin{quote}
\begin{center}
Laboratory of Theoretical Physics (LPTh) and Department of Physics,
Faculty of Sciences, University of Jijel, Bp 98 Ouled Aissa, Jijel
18000, Algeria 
\par\end{center}\end{quote}
\begin{abstract}
Adopting the thin-layer improved brick-wall method, we investigate
the thermodynamics of a black hole embedded in a spatially flat Friedmann-Lemaitre-Robertson-Walker
universe. We calculate the temperature and the entropy at every apparent
horizon for arbitrary solution of the scale factor. We show that the
temperature and entropy display a non-trivial behavior as a functions
of time. In the case of black holes immersed in universe driven by
phantom energy, we show that for specific ranges of the equation-of-state
parameter and apparent horizons the entropy is compatible with the
D-bound conjecture, even the null, dominant and strong energy conditions
are violated. In the case of accretion of phantom energy onto black
hole with small Hawking-Hayward quasi-local mass, we obtain an equation-of-state
parameter in the range $w\leq-5/3$, guaranteeing the validity of
the generalized second law.


\end{abstract}

\section{Introduction}

The discovery that the current universe is in an accelerating expansion
phase, obtained from the observations of type Ia supernovae \cite{Perlmutter-1,Riess},
inaugurate an era of intense theoretical research to understand the
mechanism driving this accelerating expansion. An immediate consequence
is that gravity behaves differently on cosmological distance scales.
A variety of possible solutions to the cosmic acceleration puzzle
have been debated during this decade including the cosmological constant,
exotic matter and energy, modified gravity, anthropic arguments, etc.
The most favored ones are dark energy based models and modified gravity
theories such $f(R)$ gravity and DGP gravity with an equation of
state parameter $w=P/\varrho<-1$ (where $\rho$ and $P$ are the
energy density and pressure of the cosmic fluid, respectively).

The dark energy component is usually described by an equation-of-state
parameter $w<-1/3$. The simplest and traditional explanation for
dark energy is a cosmological constant, which is usually interpreted
as the vacuum energy. However, the value required to explain the cosmic
expansion, which is of the order of $10^{-120},$ cannot be explained
by current particle physics. This is known as the cosmological constant
problem. On the other hand, the first year WMAP data combined with
the 2dF galaxy survey and the supernova Ia data favor the phantom
energy equation-of-state of the cosmic fluid $w<-1$ over the cosmological
constant and the quintessence field. A candidate for phantom energy
is usually a scalar field with the wrong sign for kinetic energy term
\cite{Caldwel,Nojiri}. One crucial fate of an expanding universe
driven by phantom energy is that the phantom energy density and the
scale factor diverge in finite time, ripping apart all bound systems
of the universe (galaxies, stars, atoms, nuclei), before the universe
approaches the Big-Rip singularity \cite{Caldwel,McIness}.

In this work we are interested by the effect of cosmological expansion
on local systems, namely a black hole embedded in a an expanding Friedmann-Lemaitre-Robertson-Walker
(FLRW) universe driven by phantom energy. The study of non-stationary
event horizons and their thermodynamical parameters, relevant for
a quantum theory of gravity, are currently attracting a great interest
\cite{CosmicBH}-\cite{Dyer}. In the pure expanding flat FLRW universe
dominated by phantom energy, the radius of the observer's event horizon
diminishes with time and consequently the horizon entropy, $\dot{S}_{H}<0.$
An other example is the quasi-de Sitter space, where the event horizon
and apparent horizon are different \cite{Pollock,Frolov}. It was
found that the first law and second law of thermodynamics cannot hold
when both the horizons are considered. Other authors discussed the
effect of the presence of black holes in expanding phantom energy-dominated
universe on the validity of the generalized second law of gravitational
thermodynamics (GSL) \cite{German,Lima}.

In this paper, we present a detailed calculation of the entropy of
the solution of Einstein's equations recently found and representing
a black hole embedded in an expanding Friedmann-Lemaitre-Robertson-Walker
(FLRW) universe\cite{Faraoni1,Faraoni2}. Our approach is based on
the brick-wall method (BMW) invented by t'Hooft \cite{Thooft}. In
this method a brick wall around the horizon prevents the divergence
of the free energy (or entropy) of the black hole, which is identified
with the canonical ensemble statistical-mechanical free energy (or
entropy) due to quantum excitations in thermal equilibrium with the
black hole. However, the BMW can not be used for non-equilibrium systems,
like black holes with multi-horizons, where each horizon can be considered
as an isolated thermodynamical system. The solution to the problem,
known as the thin-layer improved method, consists in considering a
thin layer near the horizon as a local equilibrium system and invoking
thermodynamics of composite systems, such that the entropy of a multi-horizons
black hole is the sum of the contributions arising from each horizon
\cite{Zhao}. Then in the improved BWM, the global equilibrium has
been replaced by a local equilibrium on microscopic scales. This method
has been applied to a wide range of black holes with multi-horizons
like Schwarzschild-de Sitter black hole, Kerr-de Sitter black hole,
Vaidya black hole, the 5D Ricci-flat black string and recently to
the calculation of the entropy in the pure expanding FLRW universe
\cite{Wontae}.

The organization of the paper is as follows: in Section II we review
the exact solution describing a black hole embedded in an expanding
Friedmann-Lemaitre-Robertson-Walker (FLRW) universe. In section III,
the statistical mechanical free energy, thermal energy and entropy
due to a scalar field are computed using the improved thin-layer BWM.
Then we discuss the conditions under which the D-bound conjecture
\cite{Bousso} is protect in expanding universe in the presence of
black holes. In section IV, our attention will be focused on the conditions
under which the second law of thermodynamics and the GSL are satisfied
when the universe is driven by phantom energy. Finally, we discuss
and summarize our results in section V.

\section{Cosmological expanding black hole}

The first solution of Einstein's theory of general relativity describing
a black hole like object embedded in an expanding universe was introduced
by McVittie in 1933 \cite{McVittie}, and is given in isotropic coordinates
by

\begin{equation}
ds^{2}=-\frac{\left(1-\frac{M_{0}}{2a(t)r}\right)^{2}}{\left(1+\frac{M_{0}}{2a(t)r}\right)^{2}}dt^{2}+a^{2}(t)\left(1+\frac{M_{0}}{2a(t)r}\right)^{4}\left(dr^{2}+r^{2}d\Omega^{2}\right),\label{eq:McVittie}\end{equation}
 where $a(t)$ is the scale factor and $M_{0}$ is the mass of the
black hole in the static case. In fact, when $a(t)=1,$ it reduces
to the Schwarzschild solution. When the mass parameter is zero, the
McVittie reduces to a spatially flat FLRW solution with the scale
factor $a(t)$. The global structure of (\ref{eq:McVittie}) has been
studied and particularly it has been shown that the solution possesses
a spacelike singularity on the 2-sphere $r=M_{0}/2,$ and cannot describe
an embedded black hole in an expanding spatially flat FLRW universe
\cite{Sussman,Nolan}.

In the following we adopt the new solution describing a black hole
embedded in a spatially flat FLRW universe \cite{Faraoni1,Faraoni2}

\begin{equation}
ds^{2}-\frac{\left(1-\frac{M_{0}}{2r}\right)^{2}}{\left(1+\frac{M_{0}}{2r}\right)^{2}}dt^{2}+a^{2}(t)\left(1+\frac{M_{0}}{2r}\right)^{4}\left(dr^{2}+r^{2}d\Omega^{2}\right),\label{eq:metric01}\end{equation}
 Using the areal radius

\begin{equation}
\widetilde{r}=r\left(1+\frac{M_{0}}{2r}\right)^{2},\quad R=a\widetilde{r},\end{equation}
 the metric takes the following suitable Painlevel$\acute{e}$-Gullstrand
form

\begin{flalign}
ds^{2}= & -\left[\left(1-\frac{2M_{0}a}{R}\right)-\frac{R^{2}H^{2}}{\left(1-\frac{2M_{0}a}{R}\right)}\right]dt^{2}+\left(1-\frac{2M_{0}a}{R}\right)^{-1}dR^{2}\label{eq:metric02}\\
 & -2RH\left(1-\frac{2M_{0}a}{R}\right)^{-1}dtdR+R^{2}d\Omega^{2},\nonumber \end{flalign}
 where $H=\dot{a}/a$ is the Hubble parameter and overdot stands for
derivative with respect to the cosmic time. We observe that the term
$R^{2}H^{2}$ plays the role of variable cosmological constant. In
order to write the metric in the Nolan gauge, we introduce the time
transformation $t\longrightarrow\bar{t}$ to remove the $dtdR$ term

\begin{equation}
d\overline{t}=F^{-1}\left(t,R\right)\left[dt+\frac{HR}{\left(1-\frac{2M_{0}a}{R}\right)^{2}-H^{2}R^{2}}dR\right],\label{eq:Time-Transform}\end{equation}
 where the integrating factor $F\left(t,R\right)$ satisfy

\begin{equation}
\partial_{R}F^{-1}=\partial_{t}\left[\frac{F^{-1}HR}{\left(1-\frac{2M_{0}a}{R}\right)^{2}-H^{2}R^{2}}\right].\label{eq:diff-exact}\end{equation}
 Then substituting $\left(\ref{eq:Time-Transform}\right)$ into $\left(\ref{eq:metric02}\right)$
and replacing $\overline{t}\longrightarrow t$ , we obtain

\begin{equation}
ds^{2}=dh^{2}+R^{2}d\Omega^{2},\label{eq:metric03}\end{equation}
 where the two-dimensional metric is

\begin{equation}
dh^{2}=-\left[\left(1-\frac{2M_{0}a}{R}\right)-\frac{R^{2}H^{2}}{\left(1-\frac{2M_{0}a}{R}\right)}\right]F^{2}dt^{2}+\left[\left(1-\frac{2M_{0}a}{R}\right)-\frac{R^{2}H^{2}}{\left(1-\frac{2M_{0}a}{R}\right)}\right]^{-1}dR^{2}.\label{eq:metric-nolan}\end{equation}
 The apparent horizons (AH) are solutions of the equation $h^{ab}\partial_{a}R\partial_{b}R=0,$
which leads to

\begin{equation}
\left(1-\frac{2m_{H}(t)}{R}\mp RH\right)\left|_{R_{A}}\right.=0,\end{equation}
 where we have used the Hawking-Hayward quasi-local mass $m_{H}(t)=M_{0}a(t).$
This mass is always increasing in an expanding universe. Discarding
the unphysical branch with the lower sign, we obtain

\begin{flalign}
R_{C}= & \frac{1}{2H}\left(1+\sqrt{1-8\dot{m}_{H}(t)}\right),\label{eq:AHC}\\
R_{B}= & \frac{1}{2H}\left(1-\sqrt{1-8\dot{m}_{H}(t)}\right),\label{eq:AHB}\end{flalign}
 where $R_{C}$ and $R_{B}$ are the cosmic and the black hole AH,
respectively. We observe that the AH coincides at a time $t_{*}$
for which $\dot{a}(t_{*})=1/8M_{0}.$ This coincidence takes place
in a future or past universe depending on the kind of matter accretion
onto the black hole.

\section{Statistical mechanical entropy}

Now we consider the statistical mechanical entropy that arises from
a minimally coupled quantum scalar field of mass $\mu$ in thermal
equilibrium at temperature $\beta$ using the thin layer improved
BWM method.

The field equation on the background given by (\ref{eq:metric02})
is

\begin{equation}
\left(\square-\mu^{2}\right)\Phi\left(x^{\mu}\right)=0,\label{eq:KG01}\end{equation}
 where $\Phi\left(x^{\mu}\right)=\Psi\left(R,t\right)Y_{lm}\left(\theta,\varphi\right)$.

Using the relation $2m_{H}(t)=R_{A}\left(1-R_{A}H\right)$, the Klein-Gordon
equation takes the form

\begin{flalign}
\left[\partial_{t}\frac{\partial_{t}+HR\partial_{R}}{1-\frac{R_{A}}{R}\left(1-R_{A}H\right)}\right.\label{eq:KG02}\\
\left.+R\partial_{R}\frac{HR\partial_{t}-\left[\left(1-\frac{R_{A}}{R}\left(1-R_{A}H\right)\right)^{2}-R^{2}H^{2}\right]}{R^{-2}\left(1-\frac{R_{A}}{R}\left(1-R_{A}H\right)\right)}+\left(\mu^{2}+R^{-2}l(l+1)\right)\right]\Psi\left(R,t\right) & =0.\nonumber \end{flalign}
In the semi-classical approximation, the wave-function is obtained
by the ansatz $\Psi\left(R,t\right)\sim e^{i\sum_{n}\hbar^{n}S_{n}(R,t)}$.
Neglecting higher orders contributions and substituting in Eq.$\left(\ref{eq:KG02}\right)$
with the ansatz $S_{0}=-i\omega t+iB(R,t)$, we obtain

\begin{flalign}
\left[\left(1-\frac{R_{A}}{R}\left(1-R_{A}H\right)\right)^{2}-R^{2}H^{2}\right]B'^{2}+2HR\omega B'+\omega^{2}\label{eq:Momentum}\\
-\left(1-\frac{R_{A}}{R}\left(1-R_{A}H\right)\right)\left(\mu^{2}+R^{-2}l(l+1)\right) & =0.\nonumber \end{flalign}
 In our derivation we have assumed the constancy of the frequency
$\omega$ near the AH and that $\dot{B}(R,t)\ll\omega$. Solving for
$B'$ we obtain

\begin{equation}
B'(R,t)\equiv p_{R}^{\pm}=-\frac{HR\omega}{f(R)\left(1-\frac{R_{A}}{R}\left(1-R_{A}H\right)\right)}\pm\frac{1}{f(R)}\sqrt{\omega^{2}-f(R)\left(\mu^{2}+R^{-2}l(l+1)\right)},\label{eq:modes}\end{equation}
 where we have set

\begin{equation}
f(R)=1-\frac{R_{A}}{R}\left(1-R_{A}H\right)-\frac{H^{2}R^{2}}{1-\frac{R_{A}}{R}\left(1-R_{A}H\right)},\label{eq-1}\end{equation}
 which is just $g_{tt}$. The sign ambiguity in Eq.$\left(\ref{eq:modes}\right)$
is related to outgoing $\left(\partial S_{0}/\partial R>0\right)$
or ingoing $\left(\partial S_{0}/\partial R<0\right)$ particles.

The number of radial modes with energy less that $\omega$ is defined
by

\begin{equation}
n\left(\omega,l\right)=\frac{1}{2\pi}\int_{R_{A}}dR(p_{R}^{+}-p_{R}^{-}),\label{n}\end{equation}
 where we used the average of the radial momentum. The second term
in (\ref{n}) is caused by a different direction such that the outgoing
and ingoing particles are taken into account. Using Eq.$\left(\ref{eq:modes}\right)$
we get

\begin{equation}
\pi n\left(\omega,l\right)=\int_{R_{A}}\frac{dR}{f(R)}\sqrt{\omega^{2}-f(R)\left(\mu^{2}+R^{-2}l(l+1)\right)}.\end{equation}
 Now, according to the canonical ensemble theory the free energy is
defined by

\begin{equation}
\beta F=\int dN\ln\left(1-e^{-\beta\omega}\right).\end{equation}
 Integrating by parts we obtain

\begin{equation}
F=-\int\frac{N\left(\omega\right)}{e^{\beta\omega}-1}d\omega,\end{equation}
 where $N\left(\omega\right)$ is the total number of modes with energy
less than $\omega$ given by

\begin{equation}
N\left(\omega\right)=\int dl(2l+1)n\left(\omega,l\right).\end{equation}
 Substituting the expression of $n$ and restricting the integration
over $l$ to the range

\begin{equation}
0\leq l\leq\frac{1}{2}\left[-1+\sqrt{1+4R^{2}\left(\frac{\omega^{2}}{f(R)}-\mu^{2}\right)}\right],\end{equation}
 the free energy takes the form

\begin{equation}
F=-\frac{2}{3\pi}\int_{\mu\sqrt{f(R)}}^{\infty}\frac{d\omega}{e^{\beta\omega}-1}\int_{R}dR\frac{R^{2}}{f^{2}(R)}\left(\omega^{2}-\mu^{2}f(R)\right)^{3/2}.\end{equation}
 The integral over the radial variable is determined by the improved
thin-layer BWM boundary conditions

\begin{alignat}{1}
\Phi\left(x^{\mu}\right)=0\textrm{ for }\textrm{ } & R_{B}+h_{B}\leq R\leq R_{B}+h_{B}+\delta_{B},\\
\Phi\left(x^{\mu}\right)=0\textrm{ }\textrm{ for } & R_{C}-h_{C}-\delta_{C}\leq R\leq R_{C}-h_{C},\end{alignat}
where $h_{B}$,$h_{C}\ll R_{B},R_{C}$ are ultraviolet cutoffs and
$\delta_{B},\,\delta_{C}$ are thickness of the thin layer, near the
BH horizon and cosmomlogical horizon, respectively. Then, the integrals
over $R$ are performed by noting that $f(R)\longrightarrow0$ in
the near vicinity of the AH. In fact $f(R)$ can be approximated by

\begin{equation}
f\left(R\right)\approx\frac{2}{R_{A}}\left(1-2R_{A}H\right)\left(R-R_{A}\right).\label{eq:expansion01}\end{equation}
 Then using the near horizon approximation $\left(\omega^{2}-\mu^{2}f(R)\right)^{3/2}\longrightarrow\omega^{3}$,
and the following integral

\begin{equation}
\int_{0}^{\infty}\frac{\omega^{3}d\omega}{e^{\beta\omega}-1}=\frac{\pi^{4}}{15\beta^{4}},\end{equation}
 the free energy is written as $F=F_{B}+F_{C}$ with

\begin{flalign}
F_{B}= & -\frac{2\pi^{3}}{45\beta_{B}^{4}}\int_{R_{B}+h_{B}}^{R_{B}+h_{B}+\delta_{B}}dR\frac{R^{2}}{f^{2}(R)},\\
F_{C}= & -\frac{2\pi^{3}}{45\beta_{C}^{4}}\int_{R_{C}-h_{C}}^{R_{c}-h_{c}-\delta_{C}}dR\frac{R^{2}}{f^{2}(R)},\end{flalign}
where $\beta_{B}$ and $\beta_{C}$ iare the Hawking temperatures
evaluated at the BH and cosmic AH, respectively. Using $\left(\ref{eq:expansion01}\right)$
we obtain

\begin{equation}
F\simeq-\frac{\pi^{3}}{90}\left[\frac{R_{B}^{4}}{\beta_{B}^{4}\left(1-2R_{B}H\right)^{2}}\frac{\delta_{B}}{h_{B}\left(h_{B}+\delta_{B}\right)}+\frac{R_{C}^{4}}{\beta_{C}^{4}\left(1-2R_{C}H\right)^{2}}\frac{\delta_{C}}{h_{C}\left(h_{C}+\delta_{C}\right)}\right].\end{equation}
 Let us now introduce the invariant distances $\widetilde{h}_{B}$
and $\widetilde{\delta}_{B}$ defined by

\begin{flalign}
\widetilde{h}_{B} & =\int_{R_{B}}^{R_{B}+h_{B}}\frac{dR}{\sqrt{f(R)}}=\sqrt{\frac{2R_{B}h_{B}}{1-2R_{B}H}},\label{eq:-1}\\
\widetilde{\delta}_{B} & =\int_{R_{B}+h_{B}}^{R_{B}+h_{B}+\delta_{B}}\frac{dR}{\sqrt{f(R)}}=\frac{\sqrt{2R_{B}\left(h_{B}+\delta_{B}\right)}-\sqrt{2R_{B}h_{B}}}{\sqrt{1-2R_{B}H}},\label{eq:Invariant BHAH}\end{flalign}
 The invariant distances near the cosmic AH, $\widetilde{h}_{C}$
and $\widetilde{\delta}_{C}$, are obtained by the substitutions $\left(h_{B},\delta_{B}\right)\longrightarrow\left(h_{C},\delta_{C}\right)$
and $\left(1-2R_{B}H\right)\longrightarrow\left(2R_{C}H-1\right)$
in the above relations. Then, the free energy takes the form

\begin{equation}
F\simeq-\frac{\pi^{3}}{45}\left[\frac{R_{B}^{5}}{\beta_{B}^{4}\left(1-2R_{B}H\right)^{3}\widetilde{h}_{B}^{2}}+\frac{R_{C}^{5}}{\beta_{C}^{4}\left(2R_{C}H-1\right)^{3}\widetilde{h}_{C}^{2}}\right].\end{equation}

The internal energy and entropy defined by $U=\frac{\partial}{\partial\beta}\left(\beta F\right)$
and $S=\beta\left(U-F\right)$, respectively, are then obtained as

\begin{flalign}
U= & \frac{\pi^{3}}{15}\left[\frac{R_{B}^{5}}{\beta_{B}^{4}\left(1-2R_{B}H\right)^{3}\widetilde{h}_{B}^{2}}+\frac{R_{C}^{5}}{\beta_{C}^{4}\left(2R_{C}H-1\right)^{3}\widetilde{h}_{C}^{2}}\right],\\
S= & \frac{2\pi^{3}}{45}\left[\frac{R_{B}^{5}}{\beta_{B}^{3}\left(1-2R_{B}H\right)^{3}\widetilde{h}_{B}^{2}}+\frac{R_{C}^{5}}{\beta_{C}^{3}\left(2R_{C}H-1\right)^{3}\widetilde{h}_{C}^{2}}\right].\end{flalign}
Using the area of the BH horizon and cosmic AH, $\mathcal{A}_{B}=4\pi R_{B}^{2}$
and $\mathcal{A}_{C}=4\pi R_{C}^{2}$, respectively, we rewrite the
entropy as

\begin{equation}
S=\frac{\pi^{2}}{90}\left[\left(\frac{R_{B}}{\beta_{B}}\right)^{3}\frac{\mathcal{A}_{B}}{\left(1-2R_{B}H\right)^{3}\widetilde{h}_{B}^{2}}+\left(\frac{R_{C}}{\beta_{C}}\right)^{3}\frac{\mathcal{A}_{C}}{\left(1-2R_{C}H\right)^{3}\widetilde{h}_{C}^{2}}\right].\label{eq:Entropy03}\end{equation}
Now the inverse temperature is calculated at the AH using $\beta^{-1}=\left|\kappa_{A}\right|/2\pi$,
where the dynamical surface gravity associated with the AH is $\kappa_{A}=\left(2\sqrt{-h}\right)^{-1}\partial_{a}\left(\sqrt{-h}h^{ab}\partial_{b}R\right)\left|_{R_{A}}\right.$\cite{Hayward},
and where the metric $h_{ab}$ is defined by $ds^{2}=h_{ab}dx^{a}dx^{b}+R(x)d\Omega^{2}$.
The calculation yields\begin{equation}
\kappa_{A}=\frac{m_{H}}{R_{A}^{2}}-H-\frac{\dot{H}}{2H},\end{equation}
 where $H=\dot{a}(t)/a(t)$, $\dot{H}=dH/dt$. We also show that the
Misner-Sharp energy defined by $M\left(R,t\right)=\frac{R}{2}\left(1-g^{\mu\nu}\partial_{\mu}R\partial_{\nu}R\right)$
is given by

\begin{equation}
M\left(R,t\right)=m_{H}(t)+\frac{HR_{A}^{2}}{2}.\label{eq:Misner-Sharp}\end{equation}
 Writing $m_{H}$ is terms of the AH, the temperature and the Misner-Sharp
energy become

\begin{equation}
\beta_{A}^{-1}=\frac{1}{4\pi R_{A}}\left|1-\frac{R_{A}}{H}\left(3H^{2}+\dot{H}\right)\right|,\label{eq:Temp}\end{equation}

\begin{equation}
M\left(R,T\right)=\frac{R_{A}}{2}.\label{eq:Misner-Sharp01}\end{equation}
 The temperature associated with dynamical horizons is not only related
to the Misner-Sharp mass on the AH, but also to the matter content
of the universe \cite{Zerbini}. Here, we point that in a series of
papers devoted to thermodynamics of expanding FLRW universe with dark
energy it is assumed the usual law $T_{A}=1/2\pi R_{A}$ for stationary
horizons\cite{Bin}, which certainly invalids the results obtained
\cite{Nouicer}. The expression of temperature (\ref{eq:Temp}) is
similar with the temperature in York theory of black holes confined
in isothermal cavities \cite{York}. In this context, a local observer
at rest will measure a local temperature which scale as $1/\sqrt{-g_{tt}}$
for a self-gravitating mass in thermal equilibrium with the boundary
of the cavity. The effect of the boundary is reflected in the relation
(\ref{eq:Misner-Sharp}), which state that the Hawking-Hayward quasi-local
mass is the sum of the thermal energy and a gravitational self-energy
due to the cavity. In our case, the black hole is embedded in an expanding
universe, and the cosmic apparent horizon plays the role of the boundary
of the cavity. On the other hand the behavior of temperature with
time is contrasted with the naive expectation that the temperature
is redshifted by the scale factor $a(t)$ relative to its value in
static background. The additional correction terms to the temperature
become important when the density energy of the universe is negligible
compared to the expansion rate. The same remark apply to the entropy
which will show a behavior with time different from the expected one
$S\sim S_{static}a^{-3}$. Here some comments are appropriate about
the assumption of local equilibrium used in the calculation, even
we are considering dynamical horizons. The improved thin-layer BMW
method is based on the fact that Hawking radiation is mainly due to
the vacuum fluctuations near the black hole horizon(s), and then allows
to identify the black hole entropy with the statistical-mechanical
entropy associated with quantum fields outside the black hole and
confined in small region(s) in the vicinity of the horizon(s). This
cure the problem of maintaining thermal equilibrium at large scales
of the original BWM of t'Hooft, without taking into account the back-reaction
of the background. However, in small regions we are dealing with the
assumption of local equilibrium and the validity of the statistical
laws we used in the calculation, and therefore need a physical justification.
The local equilibrium state is achieved if thermodynamics properties
of the system are slowly varying. The last condition when applied
to the case of the case of a black hole immersed in an expanding FLRW
universe is equivalent to the following statement: {}``local equilibrium
is achieved if local thermalization of the system is faster than the
expansion rate of the universe''. In fact assuming that $H\ll1,$
and using Eq.(39) we obtain $\frac{\delta T}{T}\sim\frac{\delta\dot{H}}{\dot{H}}-\frac{\delta H}{H}\ll1.$
This condition can be rewritten as $\frac{T}{H}\gg1,$ which is an
expression of the last statement

In the pure flat FLRW case where only the cosmic AH survives, $R_{C}=1/H$,
we obtain

\begin{equation}
\beta_{C}^{-1}=\frac{H}{2\pi}\left|1+\frac{\dot{H}}{2H^{2}}\right|\label{eq:FLRW-F}\end{equation}
 which is proportional to the de Sitter temperature, $\beta_{C}^{-1}=H/2\pi.$

Now, substituting Eq.$\left(\ref{eq:Temp}\right)$ in $\left(\ref{eq:Entropy03}\right),$
the entropy takes the form

\begin{equation}
S=\frac{1}{1440\pi}\left[\frac{\left|1-\frac{R_{B}}{H}\left(3H^{2}+\dot{H}\right)\right|^{3}}{\left(1-2HR_{B}\right)^{3}}\frac{\mathcal{A}_{B}}{4\widetilde{h}_{B}^{2}}+\frac{\left|1-\frac{R_{C}}{H}\left(3H^{2}+\dot{H}\right)\right|^{3}}{\left(2HR_{c}-1\right)^{3}}\frac{\mathcal{A}_{C}}{4\widetilde{h}_{C}^{2}}\right].\label{eq:Entropy}\end{equation}
 In order to compare our results with the standard FLRW universe,
we set the cutoffs as

\begin{equation}
\widetilde{h}_{B,C}=\sqrt{\frac{G}{180\pi}},\end{equation}
 and introduce the following factors

\begin{equation}
\mathcal{F}_{B}=\frac{\left|\frac{1}{2}-\frac{R_{B}}{2H}\left(3H^{2}+\dot{H}\right)\right|^{3}}{\left(1-2HR_{B}\right)^{3}},\quad\mathcal{F}_{C}=\frac{\left|\frac{1}{2}-\frac{R_{C}}{2H}\left(3H^{2}+\dot{H}\right)\right|^{3}}{\left(2HR_{c}-1\right)^{3}}.\end{equation}
 Then the final expressions of thermal energy and entropy take the
forms

\begin{equation}
U=\frac{3}{8G}\left(R_{B}\mathcal{F}_{B}+R_{C}\mathcal{F}_{C}\right),\end{equation}

\begin{equation}
S=\frac{1}{4G}\left(\mathcal{F}_{B}A_{B}+\mathcal{F}_{C}A_{C}\right).\end{equation}

We assume now that matter in the universe is in the form of an imperfect
fluid with a radial heat flux, described by the stress-energy tensor

\begin{equation}
T_{\mu\nu}=\left(P+\rho\right)u_{\mu}u_{\nu}+Pg_{\mu\nu}+q_{\mu}u_{\nu}+q_{\nu}u_{\mu},\end{equation}
 where $u^{\mu}=\left(\frac{A}{B},0,0,0\right)$ is the fluid four
velocity$ $, $q^{\mu}=\left(0,q,0,0\right)$ a spacial vector describing
radial heat current, and $A=\left(1+M_{0}/2r\right),\quad B=\left(1-M_{0}/2r\right).$
The general solution of the Einstein equations are given by \cite{Faraoni1},

\begin{flalign}
8\pi G\rho=3 & \left(\frac{A}{B}\right)^{2}H^{2},\label{eq:AE1}\\
8\pi Gp= & -\left(\frac{A}{B}\right)^{2}\left[3H^{2}+2\dot{H}\right].\label{eq:AE2}\end{flalign}
 Assuming $r\gg m$ and radial energy inflow ($q<0$), the accretion
rate can be written as\[
\dot{m}_{H}\simeq Ga\mathcal{A}\left|q\right|,\]
 where $\mathcal{A=\int\int}d\theta d\varphi\sqrt{g_{\Sigma}}=4\pi r^{2}a^{2}A^{4}$.
In terms of the comoving AH, these equations reduce to

\begin{flalign}
H= & \frac{8\pi G}{3}R_{A}\rho,\label{eq:E3}\\
3H+\frac{\dot{2H}}{H}= & -8\pi GR_{A}p.\label{eq:E4}\end{flalign}
 At this stage we would like to comment about the law of temperature
$T_{A}=\frac{1}{2\pi R_{A}},$ widely used in the literature. With
the aid of Eqs.(\ref{eq:E3},\ref{eq:E4}) we observe that the later
law is only valid if the energy density of the universe is significant
relative to the expansion rate

\begin{equation}
\rho\gg\dot{H}.\end{equation}
 Now, it is instructive to rewrite the dynamical surface gravity as
an invariant. In fact using the Misner-Sharp mass $M\left(R_{A},T\right)=R_{A}/2$
and defining the projection of the (3+1)-dimensional stress-energy
tensor in the normal direction of the 2-sphere, $T_{A}^{(2)}=h^{ab}T_{ab}\,\left(a,b=1,2\right)$,
we obtain

\begin{equation}
\kappa_{A}=\frac{M\left(R_{A},t\right)}{R_{A}^{2}}+2\pi R_{A}T_{A}^{(2)},\end{equation}
 where we have set $G=1$. The later invariant relation has been recently
proved in different gauges \cite{Vanzo}. On the other hand this equation
leads immediately to the analogue of first law for dynamical black
holes. In fact using $A_{A}=4\pi R_{A}^{2}$ and $V_{A}=\frac{4}{3}\pi R_{A}^{3}$
as the AH area and volume associated with the AH, we obtain

\begin{equation}
dM=\frac{\kappa_{A}}{8\pi}dA_{A}-\frac{1}{2}T_{A}^{(2)}dV_{A}.\end{equation}

Assuming now an equation of state of the form $p=w\rho$ and using
Eqs.(\ref{eq:E3},\ref{eq:E4}), we rewrite the factors $\mathcal{F}_{B,C}$
as functions of the EoS parameter $w$,

\begin{equation}
\mathcal{F}_{B}=\frac{\left|\frac{1}{2}-\frac{3}{4}x_{B}(1-w)\right|^{3}}{\left(1-2x_{b}\right)^{3}},\quad\mathcal{F}_{C}=\frac{\left|\frac{1}{2}-\frac{3}{4}x_{C}(1-w)\right|^{3}}{\left(2x_{C}-1\right)^{3}},\end{equation}
 where we defined the reduced AH $x_{B,C}=R_{B,C}H$. The factor in
the standard FLRW universe, already given in $\left(\ref{eq:FLRW-F}\right)$,
follows by setting $x_{C}=1$ in the second relation

\begin{equation}
\mathcal{F}_{0}=\left|\frac{1-3w}{4}\right|^{3},\label{eq:FLRW-w}\end{equation}
 which is exactly the entropy factor obtained in \cite{Wontae}. In
this case the temperature associated with the cosmic AH becomes

\begin{equation}
\beta_{C}^{-1}=\mathcal{F}_{0}\frac{H}{2\pi}.\end{equation}
 Since $\mathcal{F}_{0}>1$ in a universe driven by phantom energy,
this result put some doubts on the claim that the GSL in the presence
of a black hole in a phantom energy-dominated universe is protected
if the temperature is written as $\beta_{C}^{-1}=bH/2\pi,$ with $0<b<1$
\cite{Sadjadi}. In this work, the backreaction of the phantom fluid
on the black hole has been ignored as the authors assumed the usual
pure flat FLRW metric.

In figure 1 we plotted the behavior of the factors $\mathcal{F}_{B}$
and $\mathcal{F}_{C}$ as functions of the EoS parameter $w$ for
different values of the BH and cosmic AH, respectively. It is interesting
to note that the factors $\mathcal{F}_{B}$ and $\mathcal{F}_{C}$
do not depend on the AH when $w=-1/3.$

\begin{figure}[H]
 \includegraphics[width=8cm,height=8cm]{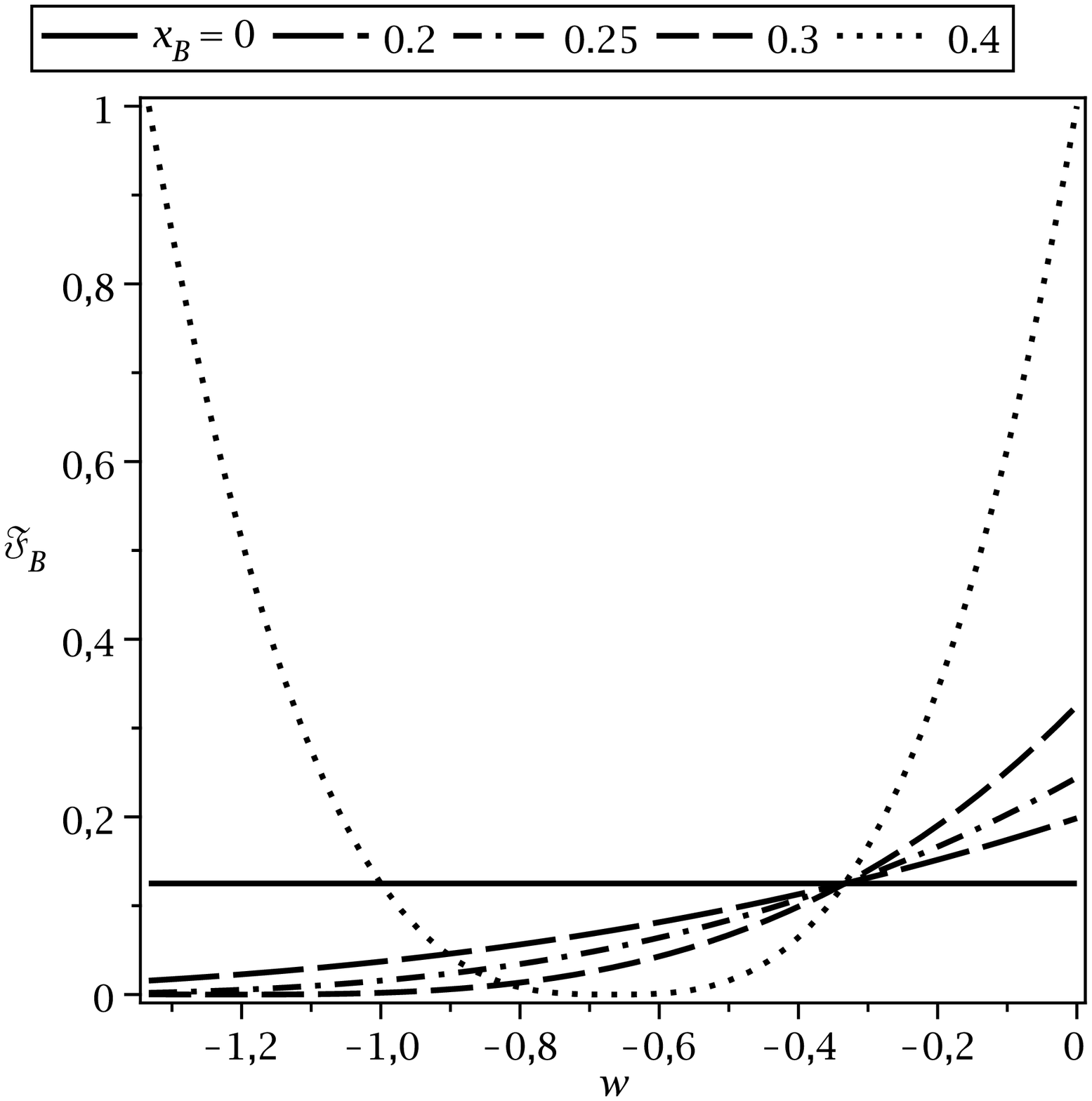} \includegraphics[width=8cm,height=8cm]{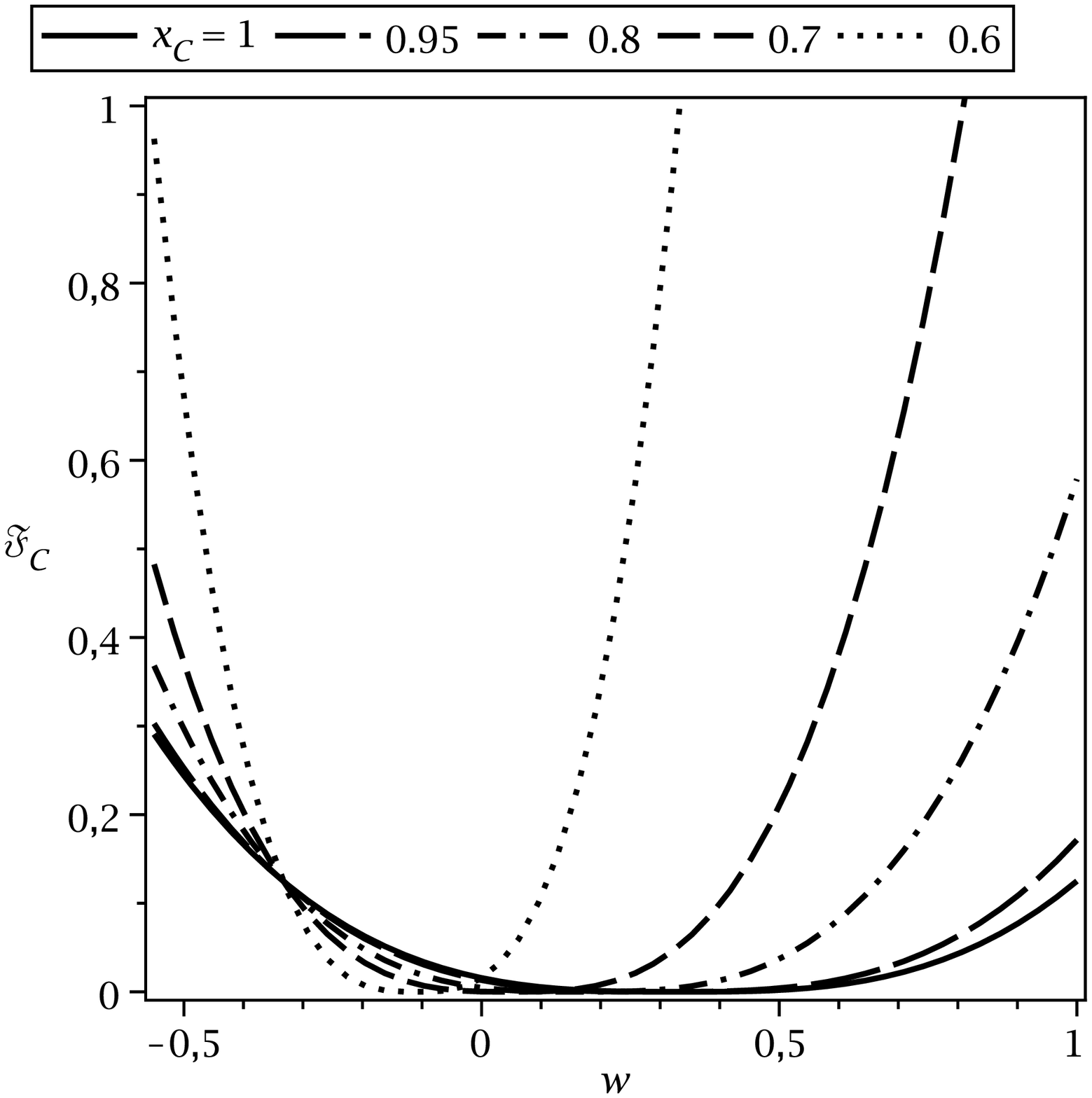}

\caption{Variation of the entropy factors $\mathcal{F}_{B}$ and $\mathcal{F}_{C}$
with the EoS parameter $w$ for different reduced AH $x_{B}$ and
$x_{C}$ on the left and right panel, respectively.}

\end{figure}

In the following let us check if the expression of the total entropy
meets the D-bound conjecture \cite{Bousso} generalized to dynamical
black holes. The later conjecture asserts that the entropy of a system
within a boundary is less than or equal to the gravitational entropy,
$S\leq S_{G}=\mathcal{A}/4G.$ We consider the special ranges of values
of $x_{B}$ and $x_{C}$ used in figure 1, and look for the possible
values of the EoS parameter $w$ for which the D-bound conjecture
is protected. In fact, from the relations giving the AH in Eqs.(\ref{eq:AHC}-\ref{eq:AHB}),
and assuming that $M_{0}\neq0$, we have $0\leq\mathcal{F}_{C}\leq1$
for $1\leq x_{C}\lesssim0.6$ and $-0.55\lesssim w\lesssim1/3$, and
$0\leq\mathcal{F}_{B}\leq1$ for $0\leq x_{B}\lesssim0.4$ and $-4/3\lesssim w\leq0$.
Then, within these ranges of the EoS parameter and the AH, it is immediate
to verify that the entropy associated with the BH and cosmic AH satisfy
the D-bound conjecture,

\begin{equation}
S_{B}\leq\frac{\pi}{4H^{2}}\;,S_{C}\leq\frac{\pi}{H^{2}},\end{equation}
 respectively. When $M_{0}=0,$ the BH horizon disappears and we obtain
$S_{C}\leq\frac{\pi}{H^{2}}$ for $-1\leq w\leq1/3.$ To verify the
D-bound conjecture for the total entropy $S_{B}+S_{C}$, we have to
verify that the following condition

\begin{equation}
\mathcal{F}_{B}R_{B}^{2}+\mathcal{F}_{C}R_{C}^{2}\leq R_{0}^{2}\label{eq:D-bound}\end{equation}
 holds, where $R_{0}=1/H$ corresponds to the AH in the pure flat
FLRW universe. Using $R_{0}=R_{B}+R_{C}$, the condition (\ref{eq:D-bound})
is rewritten as \begin{equation}
\mathcal{F}_{C}\leq1+\left(1-\mathcal{F}_{B}\right)\frac{x_{B}^{2}}{x_{C}^{2}}+2\frac{x_{B}}{x_{C}},\label{eq:DEC}\end{equation}
 where we have used the reduced AH. Since we have $x_{B}/x_{C}<1$,
the condition (\ref{eq:DEC}) becomes

\begin{equation}
\mathcal{F}_{C}+\mathcal{F}_{B}\leq4.\label{eq:cond}\end{equation}
 We easily verify that the above condition is verified for an EoS
parameter in the interval $-0.8<w<0.6$, and $0\leq x_{B}\lesssim0.4$.
and $1\leq x_{C}\lesssim0.6$. On the other hand, if we restrict the
values of the reduced cosmic AH to $1\leq x_{C}\lesssim0.8$, the
condition (\ref{eq:cond}) is satisfied for $-1.35\lesssim w<1.7.$
The later values of $x_{C}$ corresponds to a small Hawking-Hayward
quasi local mass. Finally, we point that even the phantom field is
unstable and violate the null, strong and dominant energy conditions
(Null, SEC, DEC) \cite{Cald}, the entropy of dynamical black holes
immersed in phantom energy-dominated universe, meets the D-bound conjecture.

In what follows, we would like to study the time evolution of the
thermodynamics parameters when the black hole is embedded in an FLRW
universe driven by phantom energy and accreting this cosmic fluid.
In this situation, the scale factor is given by

\begin{equation}
a_{ph}(t)=a_{0}\left(t_{rip}-t\right)^{\frac{2}{3(w+1)}},\label{eq:a1}\end{equation}
 where $t_{rip}$ in (\ref{eq:a1}) is the big rip time. In that case,
as time increases, the black hole horizon increases monotonically
while the cosmic AH decreases monotonically. The AH coincide when
$t_{*}=t_{rip}-\left[\frac{16M_{0}a_{0}}{3\left|w+1\right|}\right]^{\left|\frac{3(w+1)}{3w+1}\right|}$
, and after which the black hole singularity will become naked in
a finite time, violating the cosmic censorship conjecture \cite{Faraoni2}.
In figure 2 we show the evolution of entropy and free energy with
time for $w=-1.2$, and different initial mass of the black hole.
We observe, that entropy is decreasing and increasing at early times
and late times when approaching $t_{*}$, respectively, and diverges
at the critical AH radius $R_{C}=1/2H.$ This indicates that at times
approaching $t_{*}$, the second law of thermodynamics is protected,
$\dot{S}\geq0$, while at early times the second lwa of thermodynamics
is violated. The important result to note here is that, taking into
account the effect of backreaction of the phantom fluid on the black
hole in an expanding universe, the second law of thermodynamics becomes
valid when approaching the coincidence time, while it is always violated
in a purely phantom-energy dominated universe. 

\begin{center}
\begin{figure}[H]
 \includegraphics[width=8cm,height=8cm]{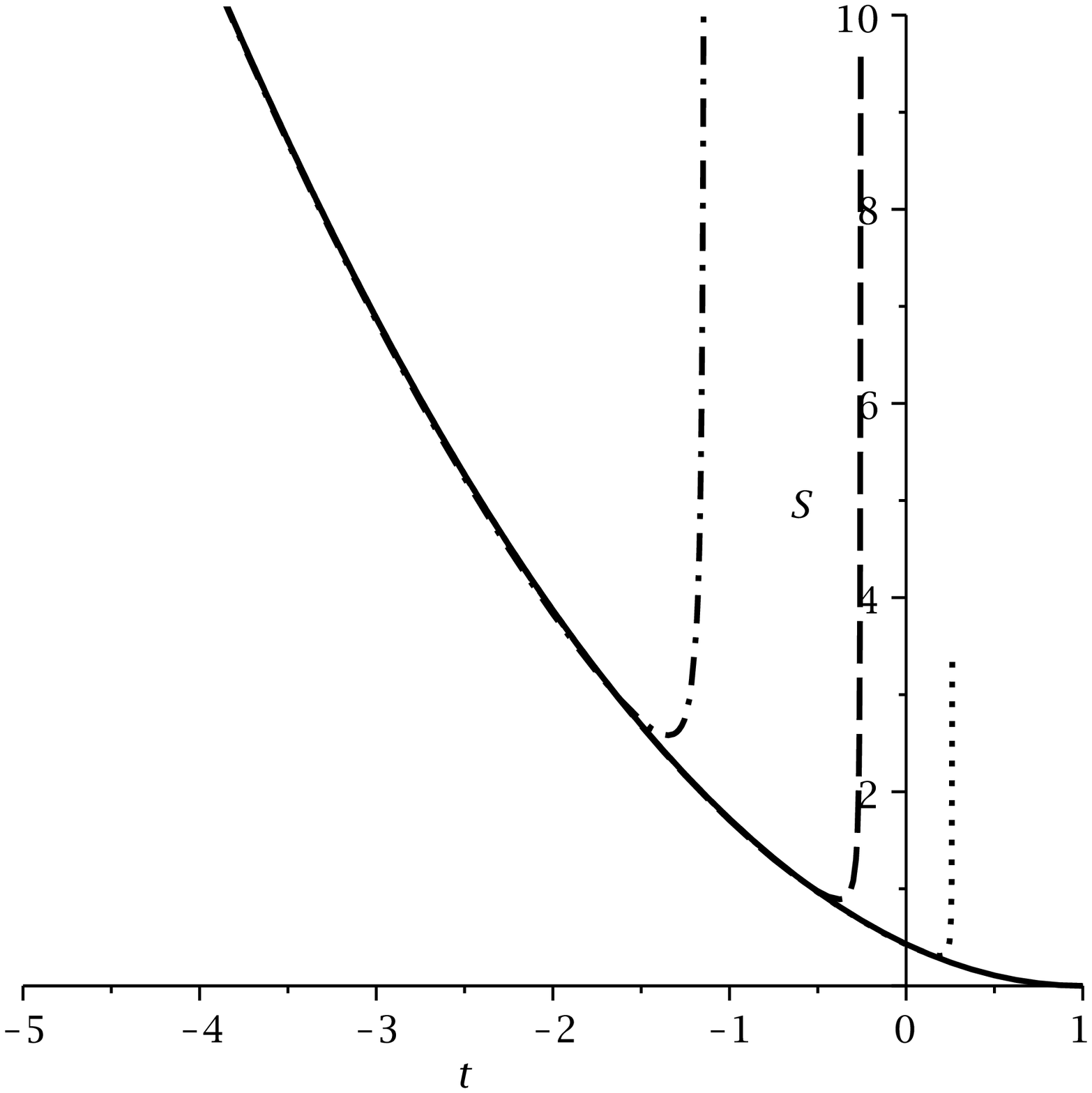} \includegraphics[width=8cm,height=8cm]{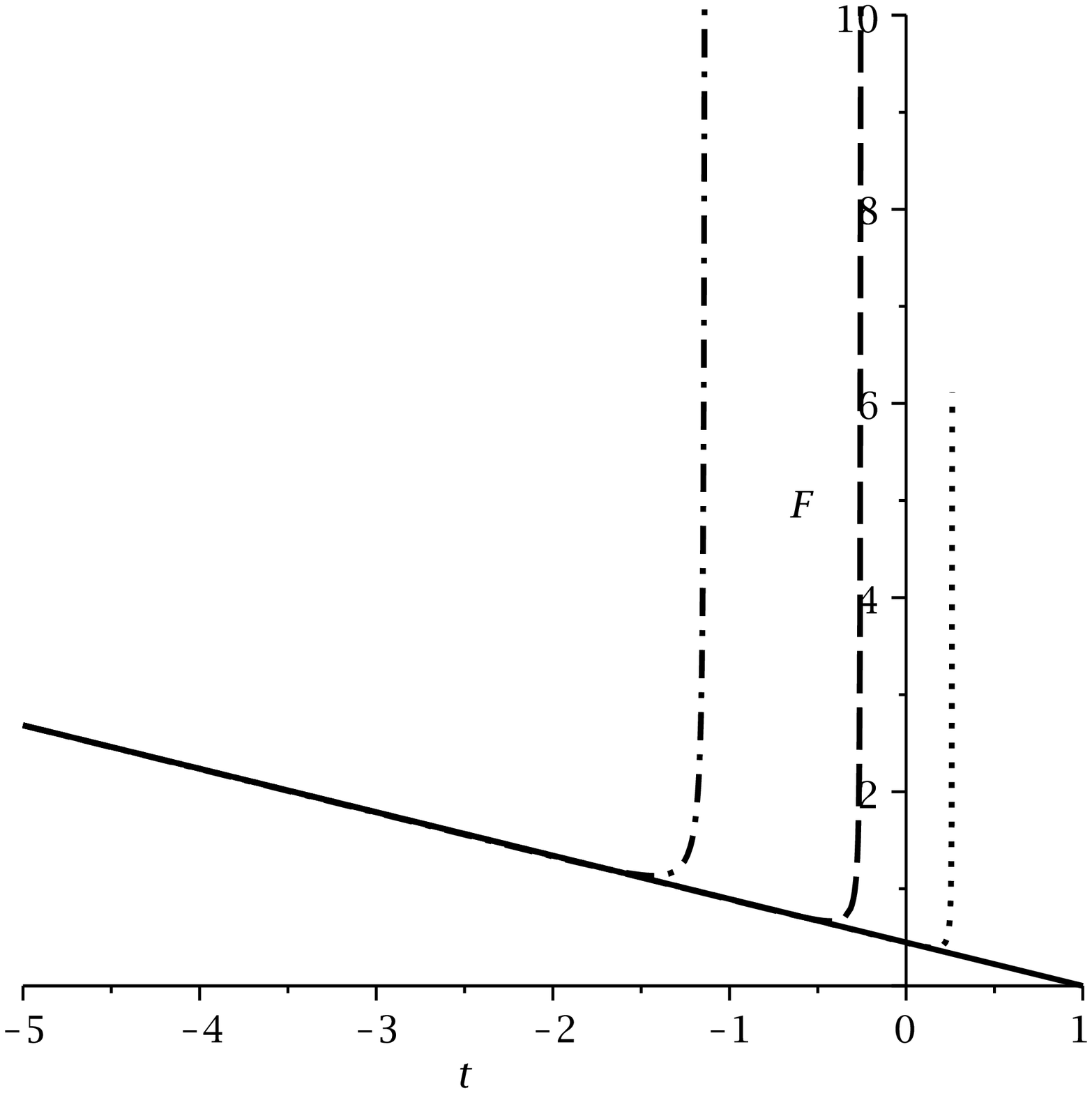}

Figure 2: Variation of entropy and free energy as functions of time
with $w=-1.2.$ From right to left we have $M_{0}=0\,\left(\textrm{solid}\right),\:0.01\,\left(\textrm{dot}\right),\:0.1\,\left(\textrm{dash}\right),\:1\,\left(\textrm{dash-dot}\right).$ 
\end{figure}

\par\end{center}

\section{Generalized second law}

It has been advanced that when a black hole is embedded in an expanding
universe driven by phantom energy, the GSL is verified under some
restrictive conditions \cite{German,Lima}. Let us now proceed to
discuss the GSL, $\dot{S}+\dot{S}_{f}\geq0$, where $S$ is the geometric
entropy associated with the AH and $S_{f}$ the entropy of the cosmic
fluid confined between the BH and the cosmic AH. The entropy of the
fluid can be obtained by using the Gibbs equation \cite{Pavon},

\begin{equation}
TdS_{f_{A}}=Vd\rho+\left(\rho+p\right)dV,\end{equation}
 where $T$ is the temperature of the energy in the vicinity of the
AH. Using the continuity equation, which is given in our model by

\begin{equation}
\dot{\rho}+\frac{\dot{R}_{A}}{R_{A}}\rho+\frac{3}{2}H\left(\rho+p\right)=0,\end{equation}
 and assuming that the cosmic fluid is in thermal equilibrium with
the boundary, the evolution of the fluid entropy is then given by

\begin{equation}
T_{A}\dot{S}_{f_{A}}=\frac{R_{A}^{2}\dot{H}}{2}\left(1-\frac{\dot{R}_{A}H}{R_{A}\dot{H}}-\frac{\dot{2R}_{A}}{R_{A}H}\right).\end{equation}
Using this relation and since the temperature is meaningful only in
the near horizon region, we assume that the fluid entropy is the sum
of the contributions near the BH and cosmic AH, respectively, 

\begin{equation}
\dot{S}_{f}=\sum_{A=B,C}\frac{R_{A}^{2}\dot{H}}{2T_{A}}\left(1-\frac{\dot{R}_{A}H}{R_{A}\dot{H}}\left[1+2\frac{\dot{H}}{H^{2}}\right]\right).\end{equation}

Let us now consider explicitly the case where the Hawking-Hayward
mass is enough small so that the quantities associated with the BH
horizon can be neglected. Then, we can set $R_{B}\backsimeq0$ and
the horizon entropy is due essentially to the contribution near the
cosmic AH

\begin{equation}
S_{C}\thickapprox\mathcal{F}_{0}\pi R_{C}^{2},\end{equation}
 where $\mathcal{F}_{0}$ is given by (\ref{eq:FLRW-w}). The time
derivative of the phantom fluid entropy becomes

\begin{equation}
\dot{S}_{f}\thickapprox\frac{R_{C}^{2}\dot{H}}{2T_{C}}\left(1-\frac{\dot{R}_{C}H}{R_{C}\dot{H}}\left[1+2\frac{\dot{H}}{H^{2}}\right]\right).\label{eq:S-phantom1}\end{equation}
 Now taking the time derivative of $S_{C}$, we obtain\begin{equation}
\dot{S}_{C}+\dot{S}_{f}=\frac{R_{C}^{2}\dot{H}}{2T_{C}}\left(1-\frac{\dot{R}_{C}H}{R_{C}\dot{H}}\left[1+2\frac{\dot{H}}{H^{2}}\right]\right)+2\mathcal{F}_{0}\pi R_{C}\dot{R}_{C}.\end{equation}
 The GSL states that $\dot{S}_{C}+\dot{S}_{f}\geq0.$ Note that in
the phantom era $\dot{H}>0$, $\dot{R}_{C}<0$, and since $T_{C}>0,$
a necessary condition for the GSL to be satisfied is\begin{equation}
1-\frac{\dot{R}_{C}H}{R_{C}\dot{H}}\left[1+2\frac{\dot{H}}{H^{2}}\right]\geq0.\end{equation}
 This condition can be integrated and leads to \begin{equation}
R_{C}(t)\leq a(t)^{\frac{3}{2}\left[\frac{1+w}{1+3w}\right]}.\end{equation}
 Using the expression of the cosmic AH, we get the following condition
on the derivative of the Hawking-Hayward quasi-local mass

\begin{equation}
\dot{m}_{H}\geq\frac{1}{8}\left[1-\left[2H(t)a(t)^{\frac{3}{2}\left[\frac{1+w}{1+3w}\right]}-1\right]^{2}\right].\end{equation}
 Now substituting the scale factor given by (\ref{eq:a1}), and the
expression of the Hubble parameter we finally obtain

\begin{equation}
\dot{m}_{H}\geq\dot{m}_{H,crit}=\frac{1}{8}\left[1-\left[\frac{4a_{0}^{\frac{3}{2}\left[\frac{1+w}{1+3w}\right]}}{3\left(1+w\right)}\left(t_{rip}-t\right)^{\frac{-3w}{1+3w}}+1\right]^{2}\right].\end{equation}
 In figure 3, we plotted the variation of $\dot{m}_{H,crit}$ with
time for different values of the EoS parameter. Knowing that the Hawking-Hayward
mass is an increasing function of time in an expanding universe, we
observe that the GSL remains protected from the past to the present
time if $w\leq-5/3.$

\begin{figure}[H]
\begin{centering}
\includegraphics[width=8cm,height=8cm]{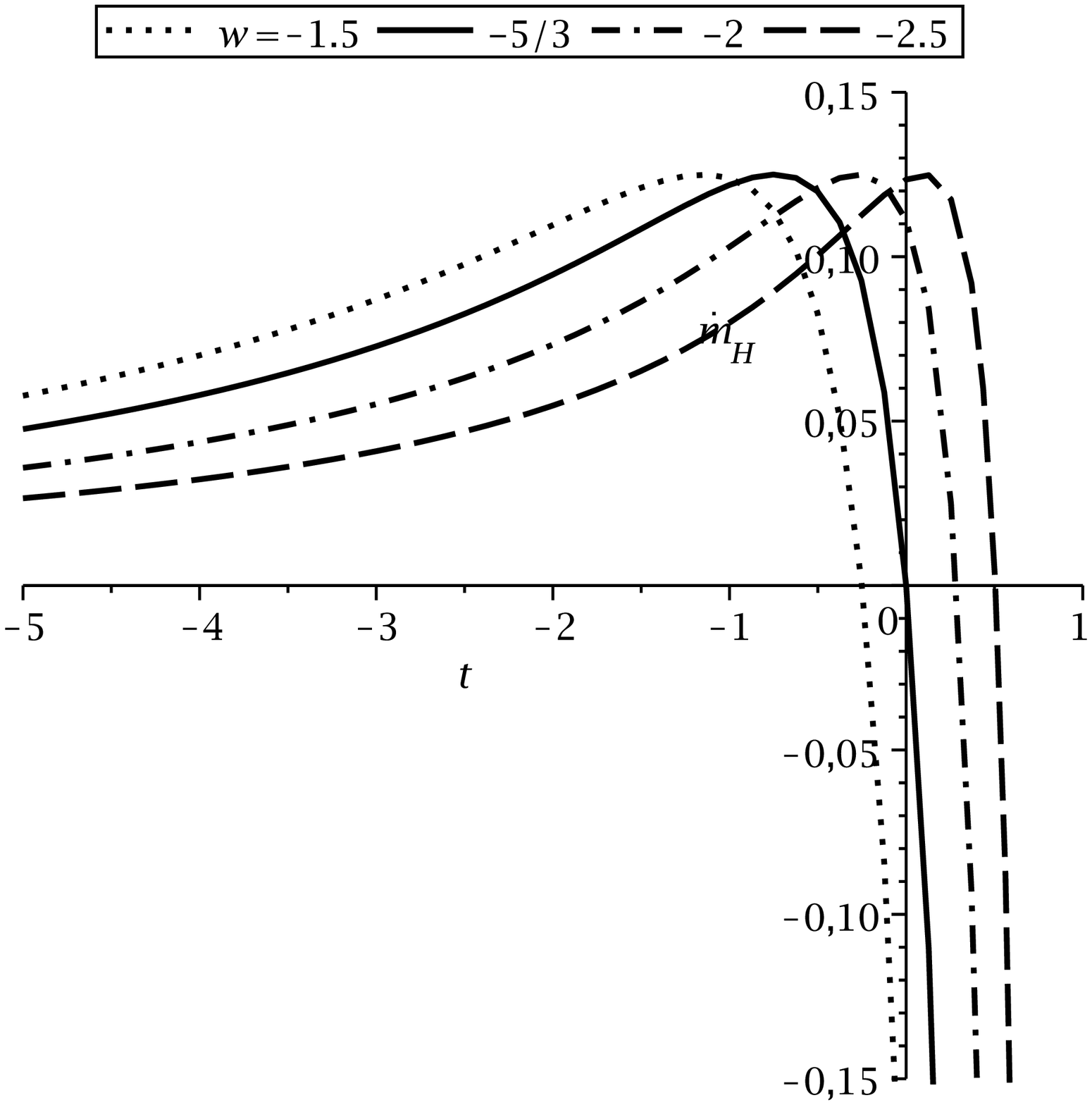} 
\par\end{centering}

\raggedright{}Figure 3: Variation of the derivative of the Hawking-Hayward
quasi-local mass as a function of time. 
\end{figure}

Using $\dot{m}_{H}=M_{0}\dot{a}(t)$ and considering the present time,
we get for the mass of the black hole the following constraint\[
M_{0}\geq-\frac{1}{9}\frac{5+3w}{\left(1+w\right)^{2}}.\]
 Assuming the positivity of the mass, we obtain again $w\leq-5/3,$
in order for the GSL to be satisfied. This value of the EoS parameter
is compatible with the analysis performed on the validity of the D-bound
conjecture in the case where the cosmic AH is close to the critical
value $1/2H.$

\[
\]

\section{Conclusion}

In this paper we have studied the thermodynamical properties of black
holes immersed in an expanding spatially flat FLRW universe. We have
particularly calculated the entropy and temperature associated with
the apparent horizons using the improved thin-layer brick wall method
and the dynamical surface gravity, respectively. The temperature and
entropy at the apparent horizons (AH) display a non trivial behavior
as a function of time, and are not scaled by the expected factors
$a(t)$ and $a^{-3}(t)$, respectively. The correction terms become
relevant when the expansion rate is significant relative to the density
energy of the universe. On the other hand, we found that the sum of
entropies associated with the AH meets the D-bound conjecture for
an EoS parameter in the interval, $-1.35<w<1.7,$ although for $w<-1,$
the null, strong and dominant energy conditions are violated. We have
also discussed the validity of the second law of thermodynamics and
the generalized second law for a black hole embedded in phantom energy-dominated
FLRW universe. The analysis showed that for arbitrary Hawking-Hayward
quasi-local mass, the second law of thermodynamics is protected when
approaching the coincidence time, at which the apparent horizons coincide,
$R_{B}=R_{C}=1/2H.$ On the other, in the case of small Hawking-Hayward
quasi-local mass, the GSL is only satisfied if $w\leq-5/3.$

\section*{Acknowledgment}

This work was supported by the Algerian Ministry of High Education
and Scientific Research for financial support under the CNEPRU project:
D01720070033.

\end{document}